\documentclass{svjour2}                    
\usepackage{graphicx}
\usepackage{amsmath}
\usepackage{amssymb}
\usepackage{bbm}
\begin{document}

\title{Evolutionary accessibility in tunably rugged fitness landscapes
}


\author{Jasper Franke \and Joachim Krug}


\institute{J. Franke \and J. Krug \at
              Institute for Theoretical Physics, University of Cologne \\
              Tel.: +49 221 470 2818\\
	      Fax: +49 221 470 5159\\
              \email{jfranke@thp.uni-koeln.de, krug@thp.uni-koeln.de}          
}

\date{Received: date / Accepted: date}

\maketitle

\begin{abstract}
The adaptive evolution of a population under the influence of
mutation and selection is strongly influenced by the structure of the underlying
fitness landscape, which encodes the
interactions between mutations at different genetic loci. 
Theoretical studies of such landscapes have been carried out for
several decades, but only recently experimental
fitness measurements encompassing all possible combinations of small sets of
mutations have become available. The empirical studies have
spawned new questions about the accessibility of optimal
genotypes under natural selection. Depending on population dynamic
parameters such as mutation rate and population size, evolutionary
accessibility can be quantified through the statistics of accessible
mutational pathways (along which fitness increases monotoncially), or
through the study of the basin of attraction of the optimal genotype
under greedy (steepest ascent) dynamics. Here we investigate these two
measures of accessibility in the framework of Kauffman's $LK$-model, a
paradigmatic family of random fitness landscapes with tunable
ruggedness. The key parameter governing the strength of genetic
interactions is the number $K$ of interaction partners of each of the
$L$ sites in the genotype sequence. In general, accessibility
increases with increasing genotype dimensionality $L$ and 
decreases with increasing number of interactions $K$. Remarkably,
however, we find that 
some measures of accessibility behave non-monotonically as a function
of $K$, indicating a special role of the most sparsely connected,
non-trivial cases $K=1$ and 2. The relation between models for fitness
landscapes and spin glasses is also addressed.

\end{abstract}

\section{Introduction and background}
\label{intro}

The basic idea of the theory of evolution as presented by Charles
Darwin \cite{Darwin} is that heritable phenotypic variability and the subsequent competition
between reproducing organisms (possibly aided by spacial separation of
habitats) shape the diversity of life as it can be observed
today. This theory, purely descriptive at first, was cast in a
quantitative form by Haldane, Fisher and Wright
\cite{h1931,w1931,f1930}. This quantification, which is
now known as the Modern 
Synthesis \cite{Huxley} of evolution founded the field of mathematical
population genetics, which has remained an active area of research ever
since. As the mutations, i.e. changes to the hereditary material of
an organism, which give rise to the diversity
in the population, cannot be controlled, they are modeled as random events
within the Modern Synthesis. Furthermore, the intricate mechanisms by
which the hereditary (or `genetic') information is processed are
accounted for in the form of stochastic models, thus methods familiar
to statistical physicists find a natural field of application in this
context. 

Within the Modern Synthesis, populations are subject to four
evolutionary forces: 
\begin{itemize}
  \item[a)] \textit{Selection} by relative average growth rate of mutant
    populations, which is the mechanism by which different
    mutants compete. Selective forces are quantified by
    \textit{fitness}, which is a measure of reproductive success and
    can be taken to be proportional to the number of offspring an
    individual leaves in the next generation.
  \item[b)] \textit{Genetic drift}, which represents random
    demographic fluctuations due to finite
      population sampling effects around the deterministic process described by
      selection alone\footnote{In the terminology of stochastic processes, this
        is the `diffusive' term in the Fokker-Planck equation
        describing the probability distribution of mutant frequencies,
        while the term describing the
        deterministic behavior is usually called `drift'. Since in
        almost all that follows, only selection will be considered,
        this should not lead to confusion.}, 
\item[c)] \textit{mutations}, which are random changes in the
  hereditary material, and   
    \item[d)] \textit{recombination}, which corresponds to two individuals
      combining (parts of) their genetic material in one common offspring.
\end{itemize} 
In this paper, we will focus specifically on the influence of selection on the
adaptive fate of a population. 

During the last decade or so, the advent of modern technologies has made  
it possible to directly manipulate or at least inspect the genetic
material of model organisms at a molecular level, thus putting
the theoretical models of adaptation that have been at the center of
attention thus far to the test. 
Of particular interest is the question of how different mutations
interact in their effects on fitness. 
To give an important example, in 
\cite{s1994} it was found that five 
mutations in the TEM-1 $\beta$-lactamase gene in the bacterium
\emph{Escherichia coli} together increase the resistance 
against a particular antibiotic by a factor on the order of
$10^5$. Besides the obvious clinical relevance of  
this finding, it raises an interesting theoretical question \cite{wwc2005,wddh2006,ch2009,fkdk2011}: Can such
a combination of mutations emerge in natural populations, where the
five mutations would have to appear and spread one at a time? In other words, is the optimal
combination of mutations \textit{accessible} to the population? In this paper,
we will address this problem in two specific regimes of adaptive
dynamics. 

\subsection{Fitness landscapes and adaptive regimes}
We start from the common assumption that all the
organism's properties are determined (in a given, fixed environment) 
by its hereditary information, which is represented as a sequence $\vec{\sigma}=\{\sigma_1, \sigma_2,
\dots, \sigma_L\}$ of length $L$ called the genotype. Real genomes
have, at the level of DNA base pairs, four different possible letters or
`alleles' at each of the $L$ sites, i.e. $\sigma_l\in \{G,A,T,C\}$. If
the proteins produced after transcription are
modeled \cite{Cell}, the number of possible alleles goes up to
$21$ (the number of amino acids). However, in many cases the number of
mutations in a population is small, such that at most two different
alleles exist at a given genomic site and the genotype can be modeled as a binary
sequence. Thus the fitness $F$ (like all other properties) is a mapping
from the Boolean hypercube to the real numbers. A natural measure of
distance between any two states is the Hamming Distance (HD) 
counting the number of point mutations separating two genotypes,
\begin{equation}
  d(\vec{\sigma}_1,
  \vec{\sigma}_2)=\sum_{l=1}^L(1-\delta_{\sigma_{1,l}, \sigma_{2,l}})
\end{equation}
where Kronecker's delta is defined by $\delta_{x,y} = 1$ 
if $x=y$ and $\delta_{x,y} = 0$ otherwise. The dynamics of a population on a given \textit{fitness landscape} (FL)
$F(\vec{\sigma})$ can be visualized as a hill climbing process, the
statistical features of which depend on the mutation rate $\mu$ per
genome, the population size $N$, and the scale $s$ of typical fitness
differences. 
In each new generation, every individual will spawn a number of
offspring that is on average proportional to that individual's
fitness. The offspring will, up to mutations, have the same genetic
configuration with the same fitness as the parent. Thus the population
in general forms a cloud in sequence space. 
An important early result of theoretical population genetics is that
the strength of demographic fluctuations ('drift') is governed by the inverse
population size $1/N$ \cite{Gillespie,Blythe2007}. 
The parameters $Ns$ and $N\mu$ therefore gauge
the strength of selection and mutations relative to the effect of
drift, giving rise to different dynamical regimes \cite{Park2010}. 
Two regimes of particular interest in this paper will be described
in the following.

\subsubsection{The strong selection/weak mutation regime}
If the product of population size and mutation rate per individual
genome is low, $N\mu \ll 1$, mutations are rare events and the
population is genetically homogeneous (`monomorphic') most of the
time. If furthermore the selection strength
is much greater than the strength of demographic fluctuations, $Ns \gg 1$,
mutations that lower fitness (called `deleterious') are rapidly driven
to extinction. This evolutionary
regime is called the `strong selection/weak mutation (SSWM)' regime
of adaptation \cite{g1983,GillespieCauses}. In this regime, the
population can only move \emph{uphill} in fitness, and only between
states of HD 1. Such a move is then called an `accessible step'. 
Between mutational steps the population is located 
at one point in sequence space.

\subsubsection{The greedy adaptive regime}
If selection is still much stronger than drift, but the mutation rate
and population size are such that $N\mu \sim L \gg 1$ and $N \mu^2 \ll
1$, all $L$ states at HD $1$ of the most populated state are generated
in one generation, but almost none at greater mutational
distance. Since selection is strong, the population will in the next
generation be dominated by the most fit of the available $L$ single
mutants and continue from that state on. Thus the population can still
be localized to a well-defined point in sequence space and the adaptive process 
takes the route of steepest fitness increase in each step, until a
(local or global) fitness maximum is reached. This regime of adaptation is
called `greedy' \cite{Gillespie,Orr2003,Jain2007,Jain2011}.  

\begin{figure}
  \includegraphics[width=0.48\textwidth]{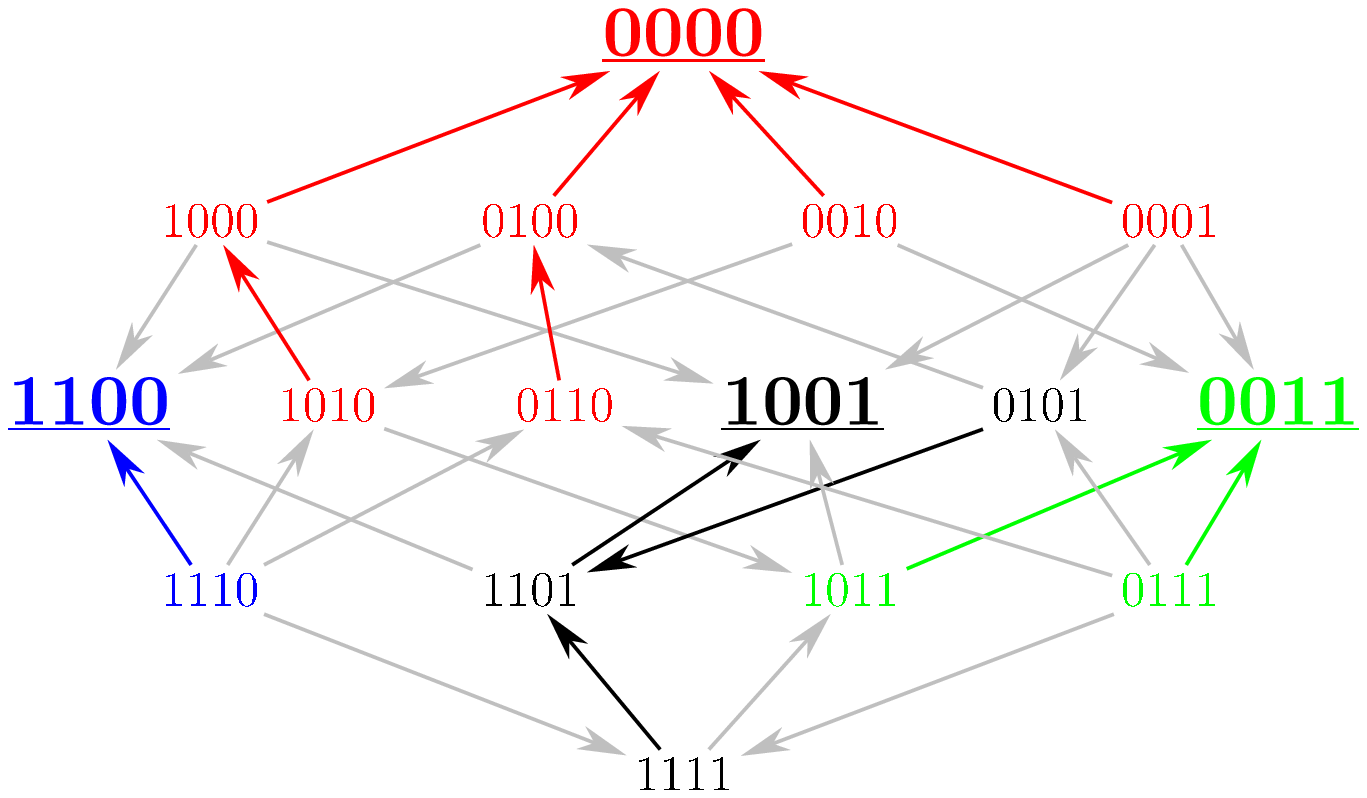}
  \hfill
  \includegraphics[width=0.48\textwidth]{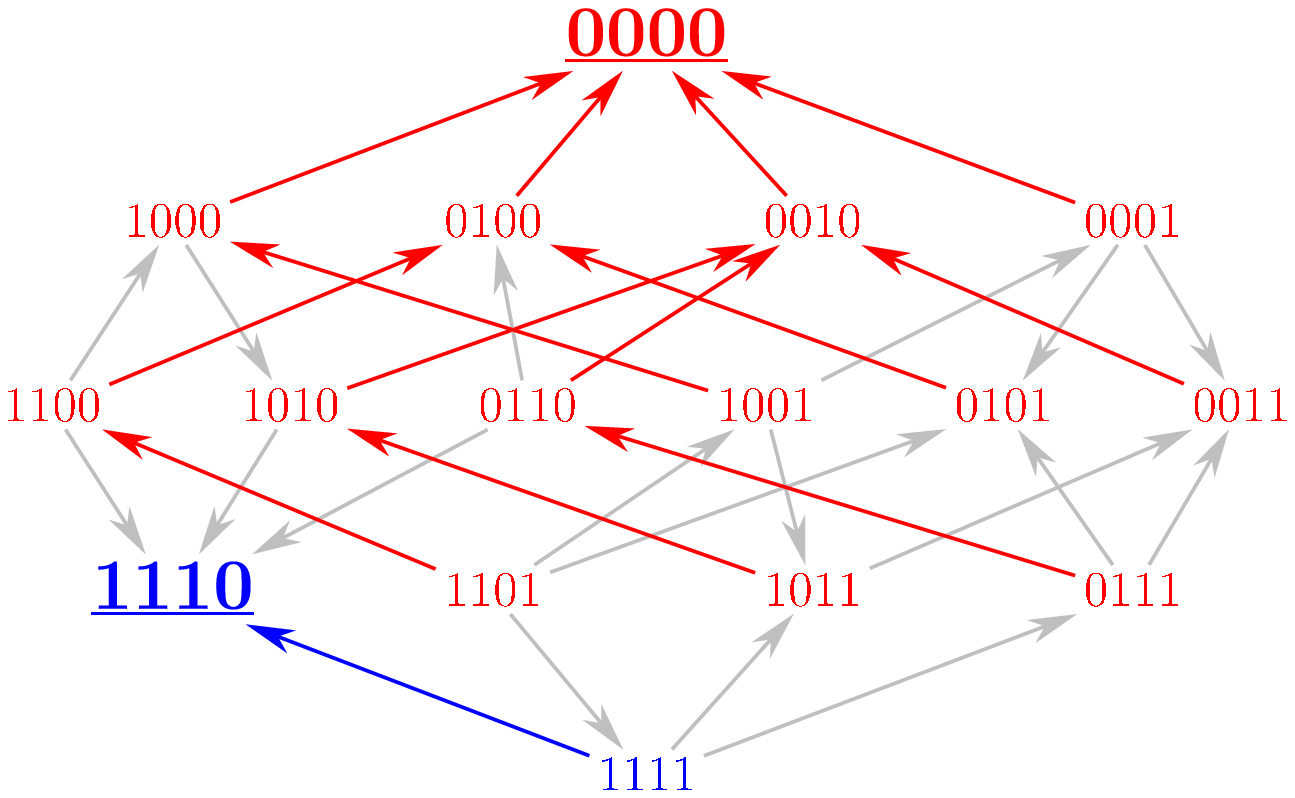}
  \caption{\label{FL_examples} (Color online) Two examples of empirical fitness landscapes
  obtained from growth rate measurements of different mutants of the 
  fungus \emph{Aspergillus niger} \cite{fkdk2011,dhv1997,deVisser2009}. The arrows between
  neighboring configurations point in the direction of fitness
  increase. Local maxima, i.e. configurations without outbound arrows
  are underlined. The colors indicate the basins of
  attraction under steepest ascent (greedy) dynamics. In the left FL
  the number of accessible paths $n$ from the antipodal genotype
  $\{1,1,1,1\}$ to the global fitness maximum $\{0,0,0,0\}$ is $n=0$ while in the right example $n=6$}
\end{figure}

\subsection{Topographic properties of fitness landscapes}

There are many ways in which the ruggedness or complexity of a fitness
landscape can be quantified. Possible measures that have been proposed
include the number of local maxima, 
the length of uphill adaptive walks,  
or the decay of fitness correlations along a random walk through the
landscape \cite{Kauffman,Weinberger1990,Stadler1999}. 
The present work is based on two particular topographic
observables which are motivated by the two adaptive regimes described
in the preceding sections. 

\subsubsection{Accessible paths} 
Within the SSWM regime, the notion of evolutionary accessibility of
the optimal genotype can be quantified in a straightforward way
\cite{wwc2005}: Since the population is constrained to move uphill in
single mutations steps, we say that a pathway connecting two genotypes
is \textit{accessible} if each mutation confers a selective advantage, i.e., if
fitness increases monotonically along the path. This definition is
closely related to the notion of reachability introduced in
\cite{Stadler2010}. Moreover, since
mutations are rare, only the \textit{shortest} (direct) pathways connecting two
genotypes will be considered\footnote{For an empirical analysis of accessible
  pathways that includes the possibility of mutational reversions see
  \cite{DePristo2007}.} . 
For two genotypes at HD $d$, there are $d!$
such pathways corresponding to the different orderings in which
the mutations separating them can occur \cite{Gokhale2009}. 

To obtain a global measure of accessibility, we focus here on pathways
to the global fitness maximum which span the entire landscape
\cite{wddh2006,fkdk2011}. 
If the global maximum is located at a sequence $\vec{\sigma}_{max}$, there is a unique
`antipodal' sequence $\vec{\sigma}_{anti}$ such that $d(\vec{\sigma}_{max},
\vec{\sigma}_{anti})=L$, and $L!$ shortest pathways connecting it to 
$\vec{\sigma}_{max}$. Our rationale for using the antipodal
sequence as a starting point is that these paths are the longest possible direct
evolutionary trajectories on the landscape, and as such their
accessibility provides a lower bound on the accessibility of typical
paths starting, for example, at a randomly chosen genotype\footnote{At
  least for the HoC and RMF models described below, it can be shown
  that the statistics of accessible paths to the global maximum does
  not depend on the choice of the starting point.}\cite{fkdk2011}. 
 On each realization of a FL, a certain number $n$ of these pathways
will be accessible in the sense defined above, see fig.
\ref{FL_examples} for two examples taken from an empirical data set
for the filamentous fungus \emph{Aspergillus niger}
\cite{fkdk2011,dhv1997,deVisser2009}. Evolutionary accessibility for an ensemble of
FL's can thus be characterized by the distribution $\pi_L(n)$,
the probability that in a given realization there will
be exactly $n$ accessible paths. 

While evolution is constrained to accessible paths only in the SSWM
regime, as a topographical feature of a FL, they can be defined
independent of any adaptive regime. In this paper, we will analyze the
behavior of the distribution $\pi_L(n)$ for a particular model of
fitness landscapes to be defined in sect.~\ref{Sec:LKmodel}. 

The study of accessible paths on FL's can be seen as a kind
of percolation problem \cite{StaufferAharony} on the Boolean
hypercube \cite{GG1997,Gavrilets}, with some important differences. 
In the (bond) percolation problem, individual
bonds between adjacent states are open or closed independently of the
others, while in the FL context, there is an effective correlation
between bonds: Since a bond on a FL is open if the fitness increases
when the bond is traversed in the direction towards the global fitness
maximum, the states of two bonds adjacent to the same sequence
are correlated via the fitness of that state. Moreover, the fact
that the accessibility of a bond is determined by the difference of a globally
defined fitness function implies a non-local gradient condition
(the arrows in fig.~\ref{FL_examples} cannot form any loops). 

\subsubsection{Basins of attraction}
Assuming a continuous range of fitness values (so that ties cannot
occur), evolutionary trajectories under 
greedy adaptive dynamics are completely deterministic and terminate at
local fitness maxima. Thus each genotype $\vec{\sigma}$ is uniquely
associated to 
a fitness maximum that will be reached by a greedy trajectory starting
at $\vec{\sigma}$. 
The set of all configurations leading to a given maximum is called the `Basin
of Attraction' (BoA) of that maximum (see fig.~\ref{FL_examples} for
illustration). Under greedy dynamics, the 
importance of a local optimum can be measured by the size of its
BoA. In particular, a landscape where the BoA of the \textit{global}
fitness maximum (GM) $\vec{\sigma}_{max}$ contains a non-vanishing
fraction of the FL for large $L$ can be said to be dominated by the GM. 

Much like accessible paths, the BoA's are of direct relevance to the
adaptation of a population only if the dynamics take place in the
greedy regime defined above, but the distribution of basin sizes can
still be considered as a 
topographical feature and analyzed independently of the adaptive
dynamics. Here, the distribution $p_L(b_{GM})$  of the size $b_{GM}$
of the BoA of the GM and its expectation value $\langle b_{GM}\rangle$ are studied in the
framework of the model of FL's to be introduced in sect.~\ref{Sec:LKmodel}.

\subsection{Fitness landscapes and spin glasses}
Models of spin glasses (SG's) were originally introduced in the
context of disordered magnets \cite{MezardParisiVirasoro} and are now
used as a paradigm for complex systems across a broad range of
disciplines,  e.g. in information theory and computational optimization
\cite{MezardMontanari,HartmannRieger}. There is an obvious close
similarity between SG's and FL's, since in both cases a real number
(energy or fitness) is associated to
each configuration of a binary (spin) chain of a given length, and the
interest is in complex landscapes with many local minima or maxima \cite{Stein1992}. 

However, the two classes of systems differ importantly in that the instantaneous
state of a spin glass is uniquely described by a single configuration, whereas a biological 
population generally occupies a cloud in genotype space. A direct
comparison between the dynamics of the two systems is therefore
meaningful only when mutations are rare, $N \mu \ll 1$, such that 
the population cloud condenses to a single point in genotype space and
adaptive steps occur one `spin flip' at a time. In this regime it can be
shown that the effective transition rates between neighboring
configurations satisfy detailed balance with respect to an
equilibrium measure of Boltzmann form, $P(\vec{\sigma}) \sim
e^{\lambda NF(\vec{\sigma})}$, with fitness $F$ taking the role of (negative)
  energy and population size $N$ acting as an inverse
  temperature\footnote{The factor $\lambda=1$ or 2 depending on
    the population dynamic model, see \cite{Sella2005}.} \cite{Berg2004,Sella2005}. 

In the spin glass context, the notion of evolutionary accessibility formulated above 
corresponds to the question of whether the ground state can be reached under a zero temperature dynamics,
where only moves that lower the system energy are accepted. Similarly, the size of the BoA of the GM
is a measure for the probability that the ground state can be reached through a steepest
descent algorithm, which is of interest in the context of optimization theory
\cite{HartmannRieger}.

\section{Fitness landscape models}
\label{Sec:LKmodel}
Even though the pace at which new empirical FL's become available has
increased greatly since the earliest examples \cite{wddh2006,dhv1997,deVisser2009}
with  several having been published in the last year alone
\cite{fkdk2011,ccdsm2011,kdslc2011,tscg2011,Szendro2012},
the number of FL's from the same system remains very
limited. Furthermore, FL's from different organisms are difficult to
compare as for example the way in which the mutations are chosen differs
strongly \cite{Szendro2012}. Thus statements about \emph{typical} behavior  
with regard to evolutionary accessibility must rely on theoretical models. 

Many such models have been proposed, each based on
different intuitions about the biological mechanisms that give rise to
fitness. For example, the `holey landscape' model assumes that mutations among
viable genotypes are effectively neutral, forming neutral networks of constant fitness 
interspersed by regions of lethality \cite{GG1997,Gavrilets}. The question of evolutionary
accessibility then reduces to the problem of percolation on the hypercube which, as
we have argued above, is rather different from the problem addressed in the present paper. 
In the following we define the two classes of fitness landscape models
that will be discussed in the remainder of this paper. 

\subsection{House of Cards models}

A very influential model is the `House of Cards' (HoC) model first
proposed by Kingman \cite{k1978,kl1987}. Based on the notion that the
genome is a finely tuned machinery and any change (mutation) to it has
the same effect as pulling one card from a house of cards, namely that
everything has to be rebuilt from scratch, the fitness values are
assumed to be independent, identically distributed random variables
(i.i.d.  RV's). This is equivalent to Derrida's random energy model (REM) of spin glasses 
\cite{d1980,d1981}.

In contrast,
the `Mt. Fuji model' \cite{Gavrilets} (named after
the famous volcano in Japan) is a model within which fitness values
are strictly ordered by HD to the GM, i.e. a state at HD $k$ from the
GM has lower fitness than any state at HD $k-1$ and greater fitness
than any state at HD $k+1$. Such a fitness landscape has smooth
flanks, no local optima and all $L!$ paths are accessible.

In terms of ruggedness, the HoC model is of maximal ruggedness with
for example the highest average number of local maxima among the
models considered here \cite{Kauffman}, while
the Mt. Fuji model is of minimal ruggedness in the same sense. Recent
studies of empirical data sets, however, imply that natural FL's are of 
\emph{intermediate} ruggedness \cite{fkdk2011,Szendro2012,Miller2011}. One model which
allows to tune the degree of ruggedness of the resulting FL is the
`Rough Mt. Fuji' (RMF) model which is a variety of the HoC model
with a superimposed linear trend towards a reference sequence
$\vec{\sigma}_0$. The fitness of a sequence $\vec{\sigma}$ 
in the RMF model is given by
\begin{equation}\label{RMF_def}
  F_\mathrm{RMF}(\vec{\sigma})=\eta_{\vec{\sigma}} -cd(\vec{\sigma}, \vec{\sigma}_0),
\end{equation}
where the $\eta_{\vec{\sigma}}$ are a family of i.i.d. RV's 
and $c>0$ is a constant \cite{fkdk2011,Aitaetal2000}. If $c \to
\infty$, the random contributions clearly play no role and the
landscape is of the Mt. Fuji type. On the other hand, for $c \equiv
0$, eq.~(\ref{RMF_def}) is simply the HoC model. The RMF model can be
seen as a HoC model (or, equivalently, REM) with a superimposed field
of strength $c$ favoring one allele in each position (that
corresponding to the reference sequence $\vec{\sigma}_0$) over the
other. 

However, the RMF model is just a phenomenological model
introduced to study the effect of intermediate ruggedness and the
`fitness field' of strength $c$ has no clear biological basis. A model where
such intermediate degrees of ruggedness arise for biologically
motivated reasons is the $LK$ model proposed by Kauffman and Weinberger\footnote{The model 
is generally known as the `$NK$ model'. Here we follow the convention of most population 
genetic literature in reserving the letter $N$ for population size and designating the
number of loci by $L$.}
\cite{Kauffman1989}. This model, which will be introduced in the next
section, is the main focus of this paper. 

\subsection{The $LK$ model}

\subsubsection{Definition}

Genes, i.e. parts of the genome that code for a protein \cite{Cell},
can be considered as the fundamental building blocks of the genetic
code. However, since the resulting proteins can only perform their
function when working together with proteins coded for by other genes,
a mutation changing the protein coded for by any given gene $A$ also
affects the function of all genes that rely on the protein produced by
gene $A$: The organism will still transcribe and translate the other
genes, thus incurring the cost of production for the corresponding
proteins, but if they cannot be put to function due to lack of the
protein produced by gene $A$, that effort is wasted. Thus if some gene
$B$ has a beneficial effect provided that gene $A$ produces its
protein, a mutation to gene $A$ may render the presence of gene $B$
deleterious, giving rise to so-called \emph{epistatic interactions}
\cite{wwc2005,deVisser2011}. 

\begin{figure}
  \includegraphics[width=0.6\textwidth]{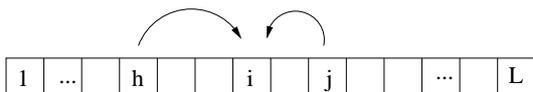}
  \caption{\label{interactions_fig} An example of the interaction
    partners of a site $i$ for $K=2$. The sites $h$ and $j$ interact
    with the site $i$} 
\end{figure}

In an attempt to capture such interactions between genes,
Kauffman and collaborators devised a model \cite{Kauffman,Kauffman1989}
within which each site $l=1, 2, \dots, L$ is assigned
a set of interaction partners  $\nu_l=\{l, \nu_{l,1}, \nu_{l,2},
\dots\nu_{l,K}\}$, consisting of indices of $K$ other
sites, see fig.~\ref{interactions_fig}; note that site $l$ itself belongs 
to the `neighborhood' $\nu_l$. In the original formulation of the
model, two ways for assigning these interaction partners were
considered: Either $K$ adjacent sites are chosen (e.g., the $K/2$ sites on
either side of $l$ or the $K$ sites following $l$), or the $K$ sites
are chosen uniformly at random, making sure that no site appears more than once
in a given neighborhood; for other possibilities see 
\cite{Stadler1999,Fontana1993,Perelson1995,Welch2005}. 
Most features of the $LK$-model appear to be rather robust with respect to the way 
in which the interaction partners are assigned. Thus in the following, only the case of randomly
chosen interaction partners will be considered.  

The fitness of a sequence $\vec{\sigma}$ is now defined as a sum over the
individual site contributions\footnote{Often the sum on the right hand side of 
(\ref{lk_fitness}) is scaled by $L^{-1}$ or $L^{-1/2}$ in order
to guarantee the existence of the limit $L \to \infty$. Here only the rank ordering 
of fitness is of interest and the overall magnitude of $F$ is irrelevant.},
\begin{equation}\label{lk_fitness}
  F_{LK}(\vec{\sigma})=\sum_{l=1}^{L}f_l(\{\sigma_j\}_{j \in \nu_{l}}).
\end{equation}
The single site contributions $f_l(\cdot)$ are taken to be
identically distributed RV's, chosen independently for each of the
$2^{K+1}$ possible arguments. 
In numerical simulations, this is realized by generating
$L$ tables containing $2^{K+1}$ i.i.d. RV's for each of the possible states of
the $L$ binary sequences of interaction partners of length $K+1$. In
the simulations reported here Gaussian RV's were used. 

The definition of the model implies that a mutation at a site $i$ of a
sequence changes the fitness contributions $f_j(\cdot)$ of all sites
$j$ that have site $i$ among their interaction
partners. This means that the case $K=L-1$, in which every site
interacts with all the others, corresponds to the HoC model: Each
mutation changes \emph{all} fitness contributions in the sum
(\ref{lk_fitness}). On the other hand, $K=0$ means that the
set of interaction partners of a given site $i$ consists only of that
site itself. Then, provided that the fitness values are drawn from a continuous
distribution, one of the two fitness values for the two states of site
$i$ will be greater than the other and 
for any suboptimal configuration, there will be point mutations that
increases fitness. Thus for $K=0$, all $L!$ paths to the (unique)
global fitenss maximum will be accessible. This limit corresponds to the Mt. Fuji model.
Any intermediate value of $K$, $0 < K < L-1$, corresponds to an intermediate
degree of ruggedness. 

\subsubsection{Relation to $p$-spin models}
 
It is a well-known fact that any fitness landscape $F(\vec{\sigma})$ defined on the
$L$-dimensional binary hypercube can be written as a linear
combination of products $\sigma_{i_1} \sigma_{i_2} \dots
\sigma_{i_p}$ with $1 \leq i_1 < \dots < i_p \leq L$ and 
$0 \leq p \leq L$ \cite{Stadler1999,Weinberger1991,Neher2011}. 
This implies that each term in the sum (\ref{lk_fitness}) can be
written as a linear combination of terms describing the interactions
among up to $K+1$ loci. Since the same argument applies to the sum as
a whole, the $LK$-model can be viewed as a superposition of $p$-spin 
spin glass models \cite{d1980,d1981} with $p \leq K+1$
\cite{Drossel2001}. However, in contrast to the fully connected $p$-spin model, where 
all possible $p$-spin interactions occur with nonzero coefficients,
the interaction graph of the $LK$-model is generally sparse, 
at least when the limit $L \to \infty$ is taken at fixed $K$ (see
below). Note, in particular, that when $K=1$ and the interaction
partners are chosen to be adjacent sites along the sequence, 
the $LK$-model is closely related to the one-dimensional Ising spin glass  
\cite{Derrida1986}. 

\section{Evolutionary accessibility in the $LK$ model}

Previous work on the $LK$-model has focused largely on local measures
of ruggedness, in particular on the density of local fitness maxima 
\cite{w1991,es2002,dl2003,lp2004} (see also \cite{Welch2005,oha2012}
for studies of adaptation dynamics on $LK$ landscapes). Here we present
results for the two  
\textit{global} measures of evolutionary
accessibility introduced above, the number of accessible
paths and the size of the basin of attraction of the global fitness
maximum. Since the biologically relevant limit
  of genome lengths on the order of millions or billions of sites is
  beyond the reach of empirical or numerical exploration, a key issue
  is to identify the asymptotic behavior of these quantities in the
  limit of large $L$. For this purpose one needs to specify how 
the number of interaction partners $K$ changes with $L$. 
We will consider three possible scalings for $K$:
\begin{itemize}
   \item[$\bullet$] $L-K$ fixed, i.e. $K \approx L$ for $L \gg 1$, 
   \item[$\bullet$] $K/L$ fixed, i.e. $K \propto L$ and 
  \item[$\bullet$] $K$ fixed.
\end{itemize}

\begin{figure}
  \includegraphics[width=0.7\textwidth]{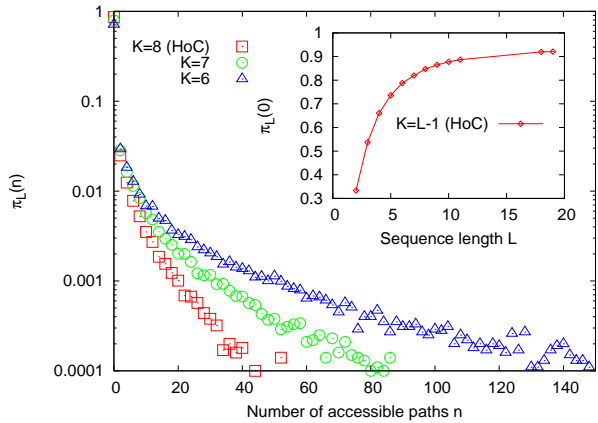}
\caption{\label{lk_paths_full} (Color online) Numerical study of the full
  distribution $\pi_L(n)$ of accessible paths for $L=9$ and
  $K=8,7,6$. Note that while for $n \geq 1$, the curves show a roughly
  exponential decay (linear in the logarithmic scales used here),
  $n=0$ carries much more weight than expected by 
  extrapolation of the exponential. The inset shows how the corresponding
  probability $\pi_L(0)$ behaves as a function of $L$ for the HoC
  model ($K=L-1$).
Unless indicated otherwise, the data shown in this and subsequent
figures is an average over $10^5$ realizations} 
\end{figure}
 
\subsection{Accessible paths}

Figure \ref{lk_paths_full} shows an example of the distribution of the
number of accessible paths, $\pi_L(n)$, for a fixed
sequence length $L=9$ and various values of $K$. One sees that the
distribution decays roughly exponentially for
large values of $n$ but has much probability weight on configurations
with no accessible paths at all. This implies that considering the
expected number of accessible paths $\langle n_L\rangle$  
does not suffice to characterize the distribution and one needs also
to study the probability $\pi_L(0)$ that in a given FL, there are no
accessible paths at all \cite{ch2009,fkdk2011}. An example for the
behavior of this probability as a function of $L$ can be seen in the inset of fig.~\ref{lk_paths_full}. 

\begin{figure}
  \includegraphics[width=0.7\textwidth]{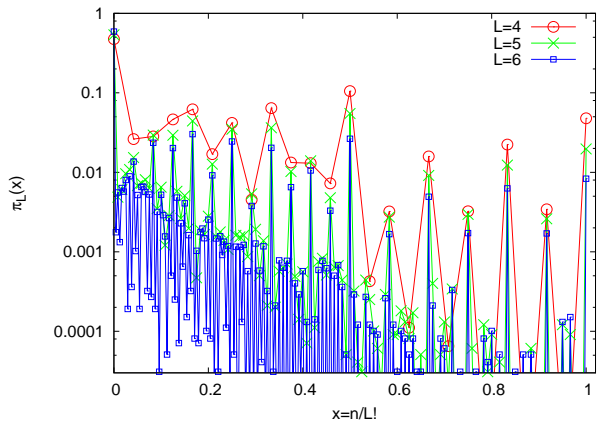}
  \caption{\label{full_K=1_fig} (Color online) Distribution of accessible paths
    for $K=1$ and several values of $L$ as a function of
    $n/L!$. This scale was chosen in order to facilitate comparison
    between different values of $L$. Note that the 
    distribution is dominated by peaks that seem to persist as $L$
    increases, with major peaks occurring at the same value of
    $x=n/L!$. In the data for $L=6$, only every third point is
    shown. This has no influence on the structure and only serves to
    render the plot less obfuscated} 
\end{figure}

\begin{figure}
  \includegraphics[width=0.48\textwidth]{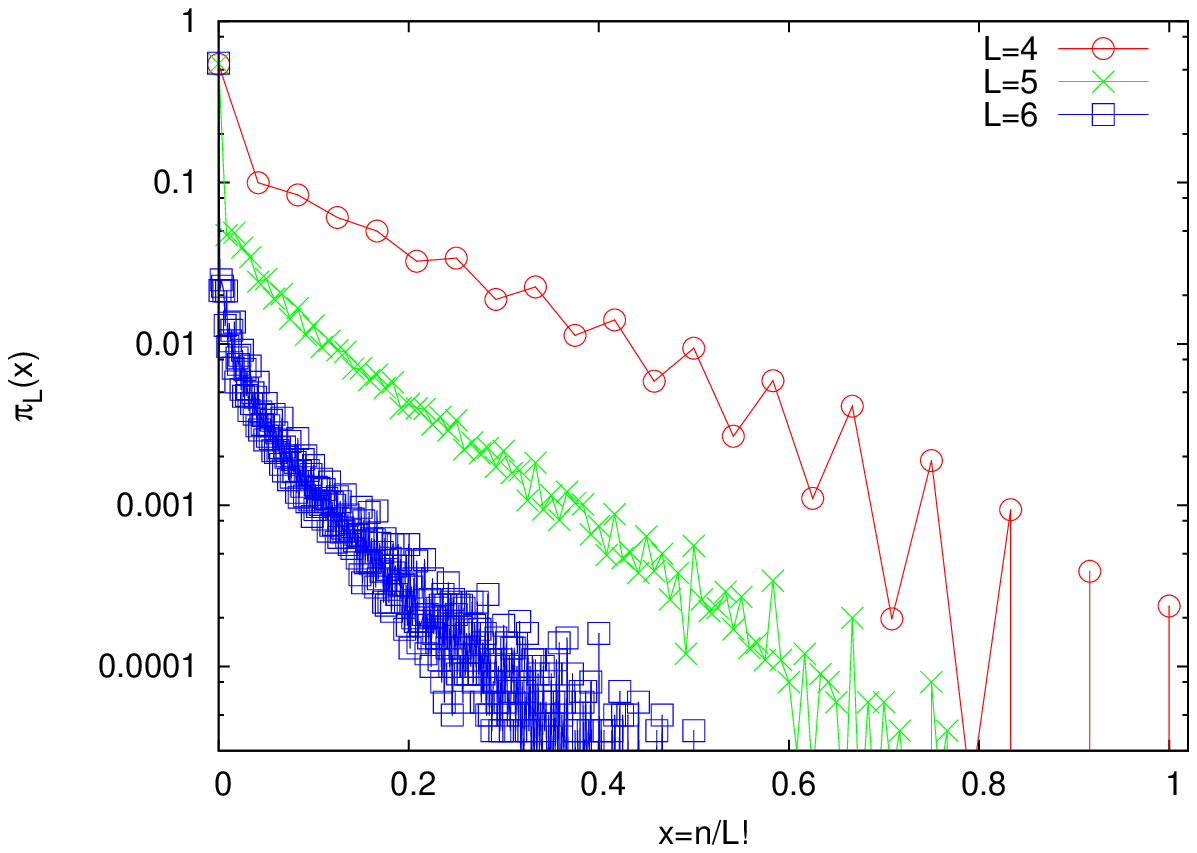}
  \hfill
  \includegraphics[width=0.48\textwidth]{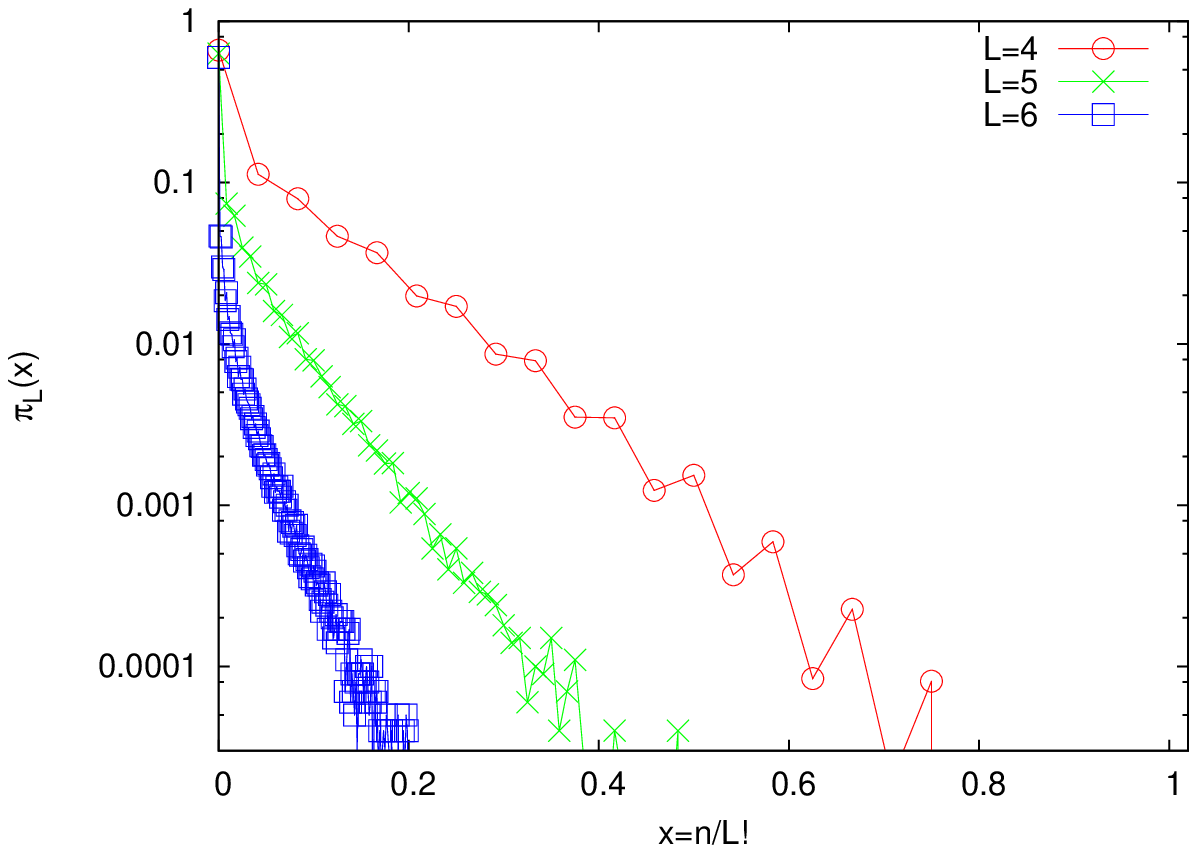}
  \caption{\label{full_K=2/3_fig} (Color online) $\pi_L(n)$ as function of $n/L!$ for
  $K=2$ (left) and $K=3$ (right) and different values of $L$. While
  the individual peaks are still discernible for small values of $L$,
  they become less pronounced as $L$ grows}
\end{figure}

\subsubsection{Full distribution for fixed $K$: Combinatorial structure}

We now consider in more detail how the shape of the distribution
$\pi_L(n)$ changes with $K$. 
For $K=0$, the distribution of accessible paths is clearly given by
$\pi_{L,K=0}(n)=\delta_{n, L!}$ since all paths are necessarily
accessible. For $K=1$, a pronounced peak at $n=L!$ is still present,
but the distribution is already dominated by realizations with no
accessible paths, see fig. \ref{full_K=1_fig}.
For $K=2$, the peak structure is still present for $L=4$ but as $L$
increases, $\pi_L(n)$ becomes smoother, see left panel of fig
\ref{full_K=2/3_fig}. This can be seen even more clearly for $K=3$,
where the curve for $L=6$ shows no peak structure at all anymore. 

The case $K=1$ seems to play a special role in this context, as the
peaks remain at the same positions (relative to the total number $L!$
of possible paths) and the peak structure is visible for all sequence
lengths for which the numerical simulations were performed. Moreover,
closer inspection of the distribution reveals that certain values of
the number of accessible paths $n$ are forbidden, in the sense that
for these values $\pi_L(n) = 0$. In particular, the peak at $n = L!$ is separated from
the rest of the distribution by a gap $\Delta n$ which increases
with increasing $L$. 

While the intriguing combinatorial structure seen in 
$\pi_L(n)$ for $K=1$ so far eludes our analytic understanding, the
computation of $\Delta n$ is rather straightforward, as it does not
depend on the details of the $LK$-model. 
To see this, we consider a fitness landscape in which all
$L!$ paths are accessible, and ask for the minimal number of pathways
that become inaccessible by reversing the fitness difference along a
single nearest-neighbor bond. The number of paths passing through a
bond connecting genotypes at HD $k$ and $k+1$ from the GM is $k!
(L-k-1)!$, which becomes minimal at $k=\lfloor L/2 \rfloor$, thus
$\Delta n = \lfloor L/2 \rfloor! (L - \lfloor L/2 \rfloor - 1)!$.
This gap is present for any value of $K$, but it is relevant only
for $K=1$, where the fully accessible state $n=L!$ still occurs with 
observable probability\footnote{Note that, in contrast to the peaks seen in
fig.~\ref{full_K=1_fig}, the gap does not live on the scale $L!$, since 
$\Delta n/L! \to 0$ for $L \to \infty$.}.  

\subsubsection{Expected number of accessible paths}

Despite the complexity of the distribution of accessible paths, some
simple analytic results are available for the expected value  $\langle
n_L\rangle$ \cite{fkdk2011,KlozerDiplom,fwk2010}. Note first that, by linearity of the
expectation, 
 \begin{equation}\label{ex_paths_eq}
  \langle n_L \rangle=\sum_{k=1}^{L!}\mathbb{P}[\mathrm{path\ } k
    \mathrm{\ accessible}]=L!\mathbb{P}[\mathrm{path\ } k
    \mathrm{\ accessible}]\equiv L! P_L 
\end{equation}
where $P_L$ is the probability for a given path to be accessible (all
paths are equivalent if one averages over realizations of
the FL). For the HoC model, $P_L$ is the probability that $L$ i.i.d. RV's (the
fitness values encountered along the path) occur in increasing order. 
There is only one such ordering among the $L!$ possible permutations,
and thus in the HoC case $P_L=1/L!$ and $\langle n_L \rangle =1$
independent of $L$. 
In the case of the RMF
model (\ref{RMF_def}) it can be shown that $P_L$ decreases
exponentially (rather than factorially) with $L$ for $c > 0$, implying
that the expected number of accessible paths grows
without bounds \cite{fwk2010}. Thus with respect to an external
field, there is a clear distinction between the maximally rugged case of
field strength $c=0$ and any positive field. 

\begin{figure}
  \includegraphics[width=0.7\textwidth]{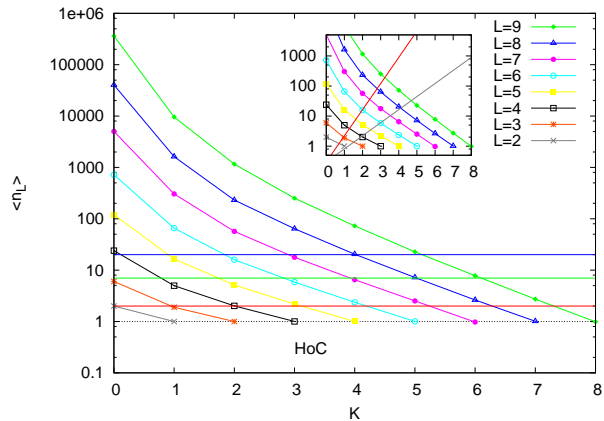}
  \caption{\label{ex_paths} (Color online) Expected number of accessible paths as function of $K$
    for various values of $L$. The horizontal lines indicate that for $L-K=2$, $3$ and $4$,
    respectively, $\langle n_L \rangle$ is almost
    constant, similar to the HoC case ($L-K=1$) where $\langle n_L
    \rangle = 1$ exactly. In the inset, the
    points for $L/K=2$ (lower line) and $L/K=3$ (upper line)
    are connected by functions of the form $0.3 e^{\beta K}$ with
    $\beta=1$ for the lower and $\beta=2$ for the upper line. This
    indicates that the average number of accessible paths grows
    exponentially with $L$ when $L/K$ is kept fixed}
\end{figure}

A similar distinction is observed numerically between
the $L$-dependence of $\langle n_L\rangle$ when $L-K$ is kept fixed 
and the other two scalings, as can be seen in fig.~\ref{ex_paths}. In this plot, the expected number of paths is depicted as
a function of $K$ for various values of $L$. The last point for each
value of $L$ corresponds to $K=L-1$. These points all lie on the line $\langle
n_L \rangle=1$, confirming the HoC result. The second to last points
correspond to $L-K=2$ and so on. While the data for $L-K \geq 2$ do
not lie strictly on a line of constant $\langle n_L \rangle$, they
increase very slowly with $K$. On the other hand, data points
belonging to a fixed value of $K$ are almost equidistant along the ordinate in the semi-logarithmic
plot, implying exponential growth. Similarly, 
if the ratio $L/K$ is kept fixed (for example at $L/K=2$ or
$L/K = 3$, see inset of fig. \ref{ex_paths}) a roughly exponential or
slightly super-exponential growth of  $\langle n_L \rangle$ with $L$ is observed.

\subsubsection{Probability of finding no accessible path}

As was mentioned above in connection with the first numerical example of
$\pi_L(n)$ in fig. \ref{lk_paths_full}, in addition to the expected
number of accessible paths it is useful to consider also the
probability $\pi_L(0)$ that there is no accessible path. 
The complementary probability $1-\pi_L(0)$ is the
probability that there is at least one accessible path to the global
fitness maximum, which can be viewed as an overall measure of
accessibility \cite{ch2009,fkdk2011}. 

For the RMF model defined in eq. (\ref{RMF_def}), numerical
simulations show that (similar to the expected number of accessible
paths) the cases $c=0$ and $c >0$ behave very
differently for large $L$ \cite{fkdk2011}. In particular, 
for $c>0$, $\pi_L(0)$ is a non-monotonic function of $L$ 
which declines for large $L$, indicating increasing accessibility, while
for the HoC case $c=0$, $\pi_L(0)$ increases monotonically with $L$
and appears to approach the limit $\lim_{L \to \infty} \pi_L(0) =1$. 

\begin{figure}
  \includegraphics[width=0.48\textwidth]{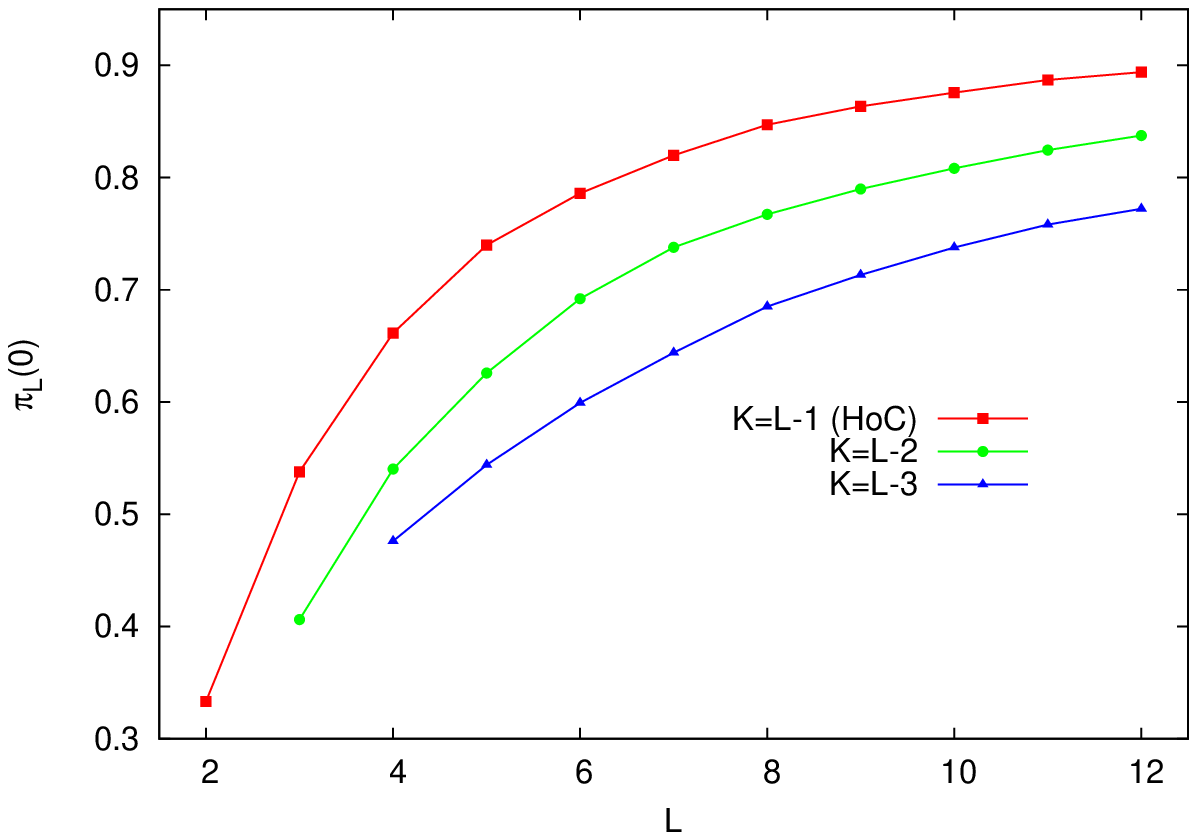}
  \hfill
  \includegraphics[width=0.48\textwidth]{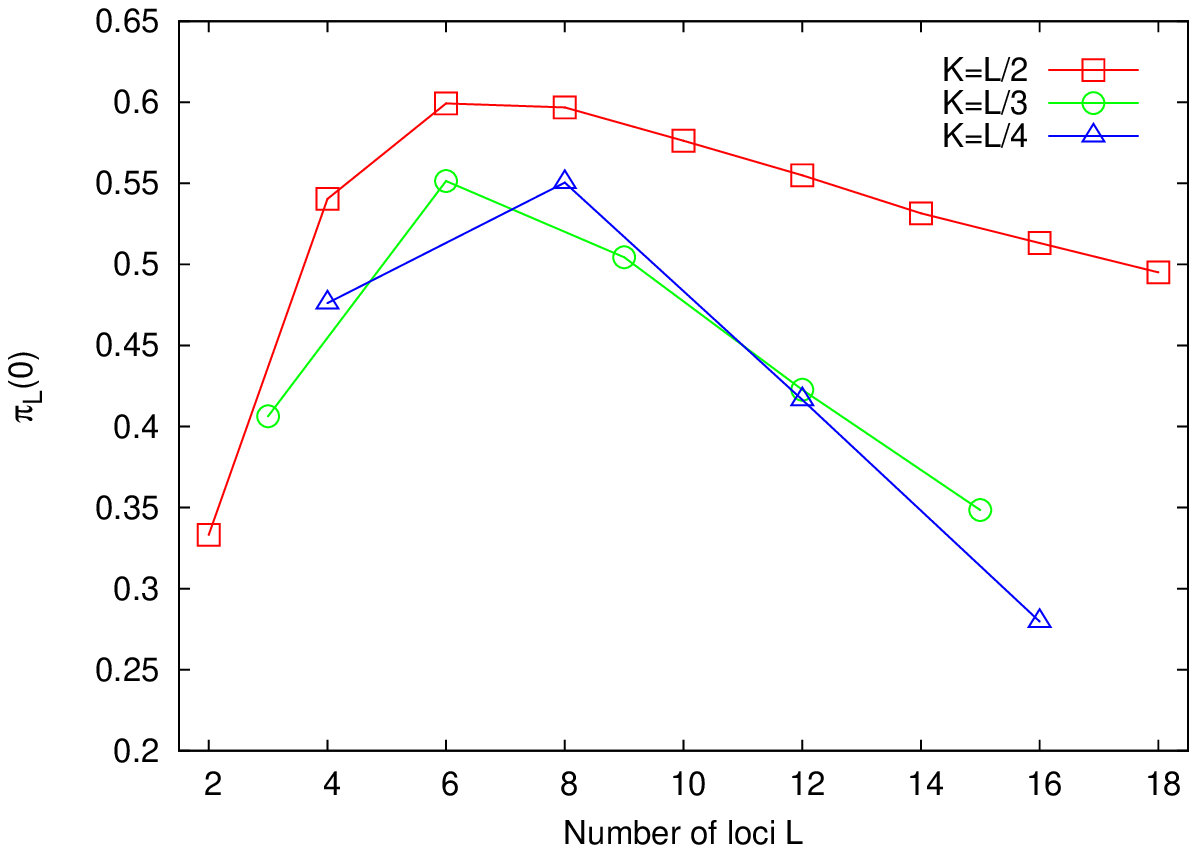}
  \caption{\label{p0_dif_fix} (Color online) Left: The probability $\pi_L(0)$ of no
    accessible path as a function of $L$, keeping $L-K$ fixed. The
    curve $K=L-1$ corresponds to the HoC case. All three curves
    increase for the entire range of $L$-values explored (up to
    $L=20$ for the HoC case, not shown here). Right: $\pi_L(0)$ at
    fixed ratio $K/L$ }
\end{figure}

The left panel of fig. \ref{p0_dif_fix} shows the behavior of
$\pi_L(0)$ for the $LK$-model at fixed values of $L-K$.  The curves 
increase monotonically in a very similar fashion. The curve for
$K=L-1$ represents the HoC case which was studied in more detail in
\cite{fkdk2011}, where simulations showed that it remains monotonic at least
up to $L=20$. The
right panel of fig.~\ref{p0_dif_fix} shows the behavior of $\pi_L(0)$
if the fraction of interaction partners $K/L$ is kept fixed. These
curves show  non-monotonic behavior with an increase up to a certain
value of $L$ and a subsequent decrease. Thus the simulations indicate
that for large $L$, there will with high probability (possibly with
probability $1$) be accessible paths in such a FL.

The results shown so far suggest a relatively simple picture, which
matches that found previously for the RMF model
\cite{fkdk2011}. Increasing $L$ at fixed $L-K$ one observes HoC-like
behavior, with accessibility (measured by $1-\pi_L(0)$) decreasing
monotonically with $L$. On the other hand, when the fraction of
interacting sites is kept fixed, accessibility increases with $L$, 
and decreases (for large enough $L$) with increasing $K/L$, indicating
that the parameter $K/L$ plays a role similar to the field $c$ in the RMF model. 

\begin{figure}
  \includegraphics[width=0.6\textwidth]{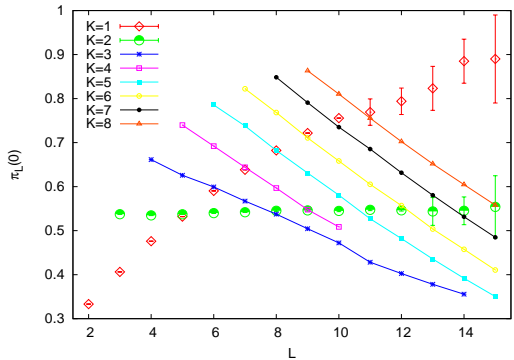}
  \caption{\label{p0_kfix} The probability $\pi_L(0)$ that a given
    realization of a FL has no accessible paths at all as function of
    the sequence length. Note that while the curves for $K\geq 3$
    decrease with $L$, the curve for $K=1$ actually increases for the
    range of $L$ considered here. The error bars are due to a lower
    number of samples for larger systems ($10^3$ or $10^4$)}
\end{figure}

However, the data depicted in fig. \ref{p0_kfix} for the behavior of
$\pi_L(0)$ at fixed $K$ reveal a more complex scenario.  
The results for $K\geq 5$, which were already reported in
\cite{fkdk2011},
show that, as expected, $\pi_L(0)$ decreases monotonically with
$L$, a behavior that is seen to extend also to $K=4$ and $K=3$. 
In striking contrast, the curve for $K=2$ seems to be almost
independent of $L$, 
and even more surprisingly, for $K=1$ $\pi_L(0)$ \emph{increases} with
$L$. This non-monotonic $K$-dependence of the accessibility measure 
$1-\pi_L(0)$ is highly unexpected and cannot be inferred, e.g., from
the results for the expected number of accessible paths. Together with
the characteristic differences in the shape of the distribution
$\pi_L(n)$ (see above) and the corresponding distribution of basin
sizes (see below), the results shown in fig. \ref{p0_kfix} provide a
strong indication for a qualitative change in the $LK$ fitness
landscapes at $K=2$. We will return to the possible reasons for this
puzzling behavior below in sect.~\ref{sec:Conclusions}.  

\subsection{Basin of attraction of the global maximum} 

Under greedy dynamics, the configuration space can be decomposed into
disjoint BoA's, each belonging to one specific local optimum. Thus the
expected size of a \emph{typical} BoA, $\langle b_{typ} \rangle$,
can be estimated by dividing the total number of states by the
expected number of optima $\langle N_{opt}\rangle$ \cite{w1991}. For the HoC model,
the average number of local optima can easily be computed: Since a
local optimum has to have a greater fitness than each of its $L$
neighbors [which happens with probability $1/(L+1)$], the expected number
of local optima is $\langle N_{opt}\rangle=2^L/(L+1)$ and
hence\footnote{Note that this estimate relies on uniformly sampling
  all local maxima. The size of a typical basin found by a greedy walk
starting at a random position will be larger, because this
introduces a bias towards larger basins.}
\begin{equation}
  \langle b_{typ}\rangle =\frac{2^L}{\langle N_{opt} \rangle}=L+1.
\end{equation}
Since asymptotic results for the number of local maxima in the $LK$
model exist \cite{w1991,es2002,dl2003,lp2004},
one can straightforwardly obtain analogous estimates for $\langle
b_{typ} \rangle$ in this case. However, this only gives a lower bound
to the size of $\langle b_{GM}\rangle$ of the BoA of the GM which grossly underestimates 
the actual behavior and will not be presented here.

\begin{figure}
  \includegraphics[width=0.7\textwidth]{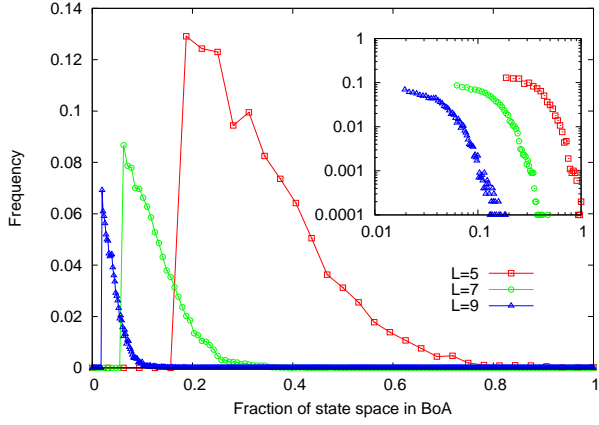}
\caption{\label{HoC_BoA_fig}  (Color online) Distribution of the size of the basin of
  attraction of the GM for the HoC model as function of the rescaled
  basin size $b/2^L$. As the sequence
  length $L$ increases, the distribution becomes more concentrated near
  the minimum value $b_{min} =  L+1$. Inset shows the same data in
  double-logarithmic scales}
\end{figure}

One reason for the difference in behavior between $\langle b_{typ} \rangle$
and $\langle b_{GM}\rangle$ is that a typical, i.e. only local
maximum has to compete for each of its nearest neighbor 
states with another local optimum,
whereas the GM does not have to compete for its nearest
neighbors: By definition, for each neighboring sequence the GM is the
state of highest fitness in \textit{its} neighborhood. 
Thus the BoA of the GM comprises at least $L+1$
configurations, whereas the BoA's of typical maxima can be (and often
are) smaller, see fig.~\ref{FL_examples} for illustration. Numerical simulations of the distribution  $p_L(b_{GM}$
of the BoA for the HoC model show that as $L$ grows, this distribution
becomes more concentrated around the minimal value $L+1$, see fig.
\ref{HoC_BoA_fig}. Since the HoC model is maximally rugged, the question
remains whether the BoA of the GM comprises a finite fraction of the
entire state space for the $LK$-model with $K < L-1$. 
Before addressing this question, the full size distribution of the
BoA of the GM will be studied.

\begin{figure}
  \includegraphics[width=0.7\textwidth]{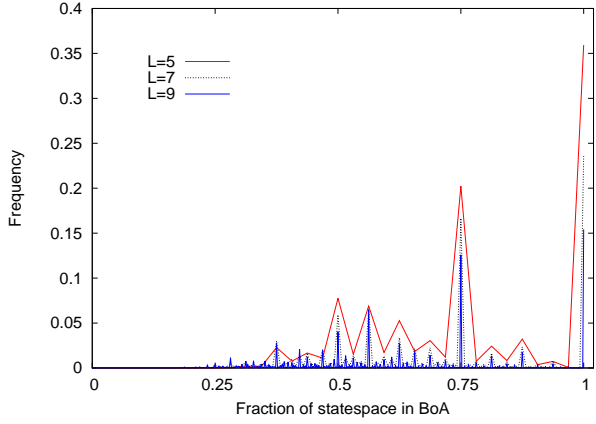}
  \caption{\label{fd_BoA_K1} (Color online) Size distribution of the GM's BoA
    for $K=1$ and different values of $L$ as function of the fraction
    of state space $b/2^L$. Note that the spikes
    dominating the distribution remain visible and occur at the same
    fraction of $b/2^L$ for all values of sequence length considered}
\end{figure}

\begin{figure}
  \includegraphics[width=0.48\textwidth]{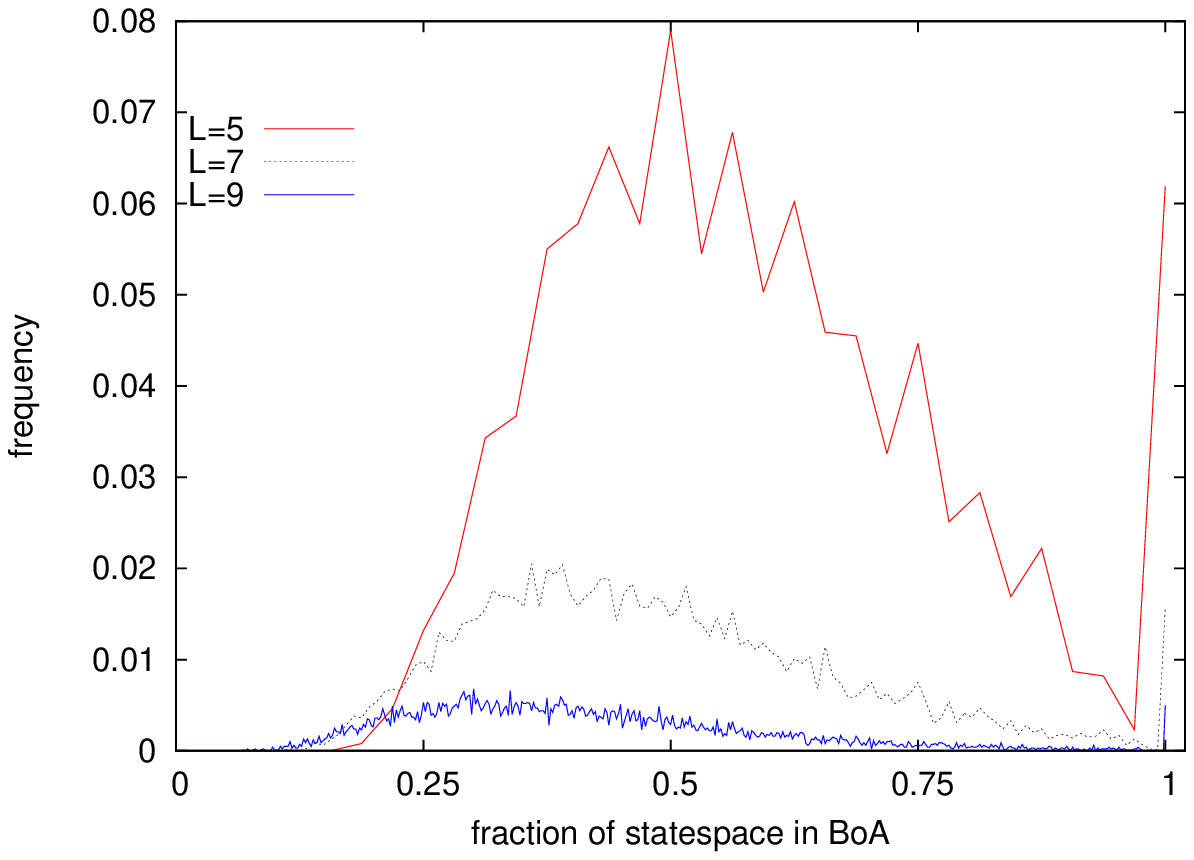}
  \hfill
  \includegraphics[width=0.48\textwidth]{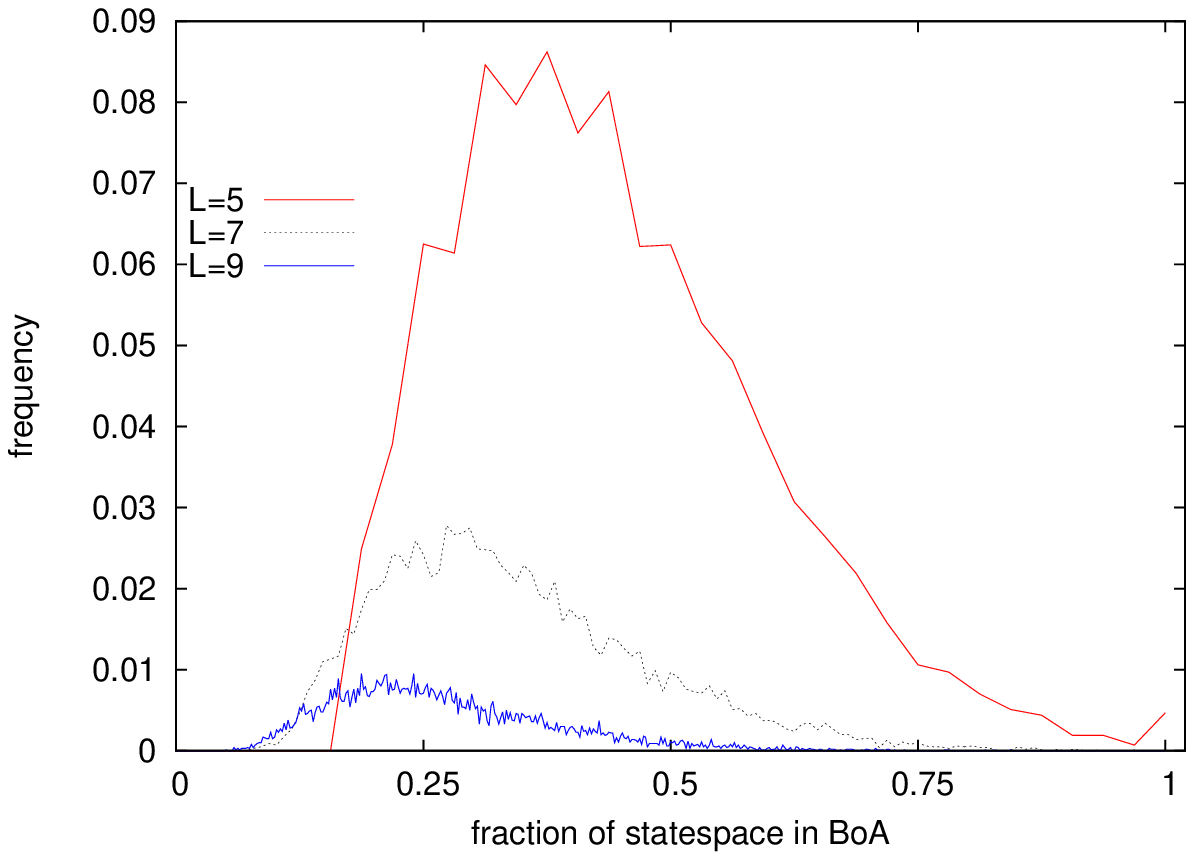}
  \caption{\label{fd_BoA_K2/3} (Color online) Size distribution of the BoA of the GM
    for $K=2$ (left) 
    and $K=3$ (right) for various sequence lengths as function of the
    fraction of sequence space. Note that unlike the case $K=1$, here the
    discrete structures vanishes as $L$ grows} 
\end{figure}

For $K=1$, the distribution of the GM's BoA shows a striking discrete structure
with spikes that remain at fixed values of $b/2^L$ as $L$ grows. As can be seen in fig.
\ref{fd_BoA_K1}, the most important contribution comes from
realizations in which the entire state space is in the BoA, followed by
contributions from realizations with exactly $3/4$ of the state space in
the BoA. Similar to the distribution of the number of
accessible paths for $K=1$ shown in fig. \ref{full_K=1_fig}, the
fact that the positions of dominant peaks remain at the same fraction
of state space indicates that the effect causing these `spectra' is commensurate
(in some sense) with the total size of sequence space. 
Some amount of discrete structure remains visible for $K=2$ and $3$,
but as can be seen in fig.~\ref{fd_BoA_K2/3}, this appears to be a finite-size effect
which becomes less prononouced as $L$ increases.

\begin{figure}
  \includegraphics[width=0.48\textwidth]{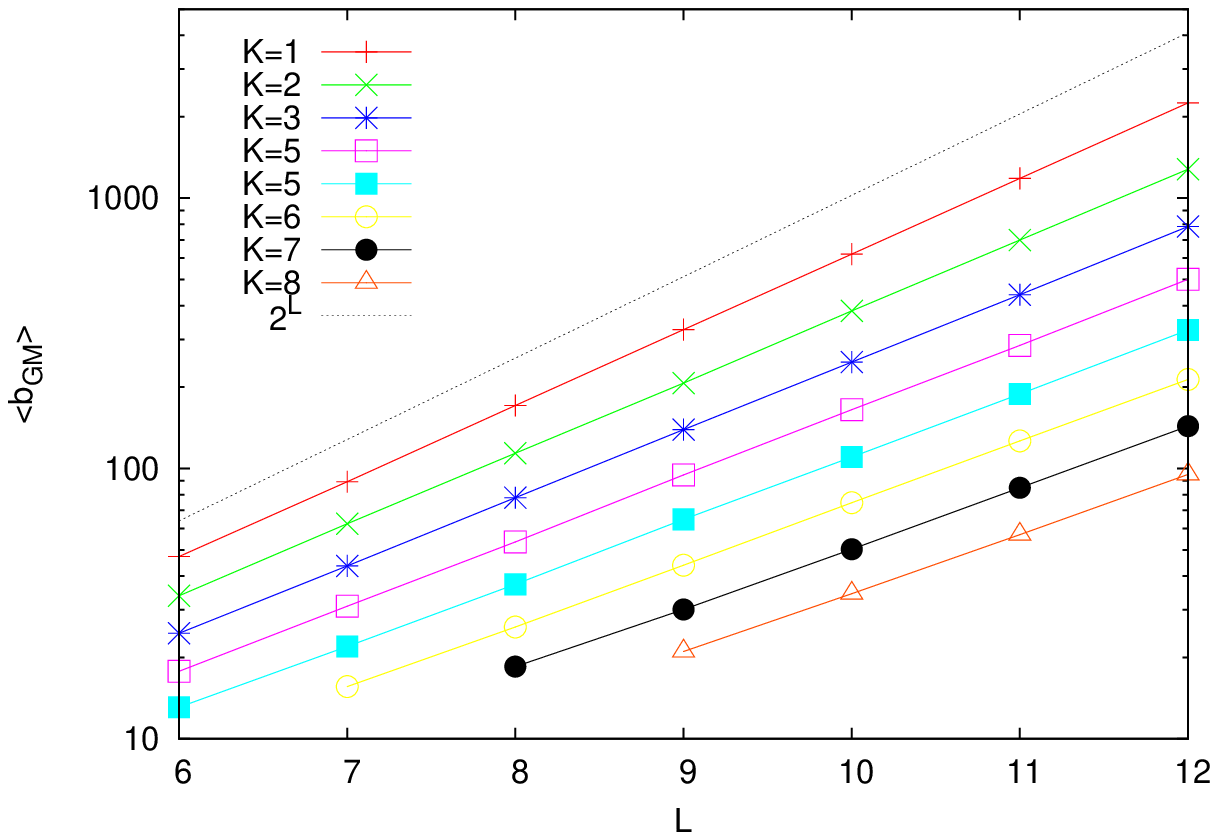}
  \hfill
  \includegraphics[width=0.48\textwidth]{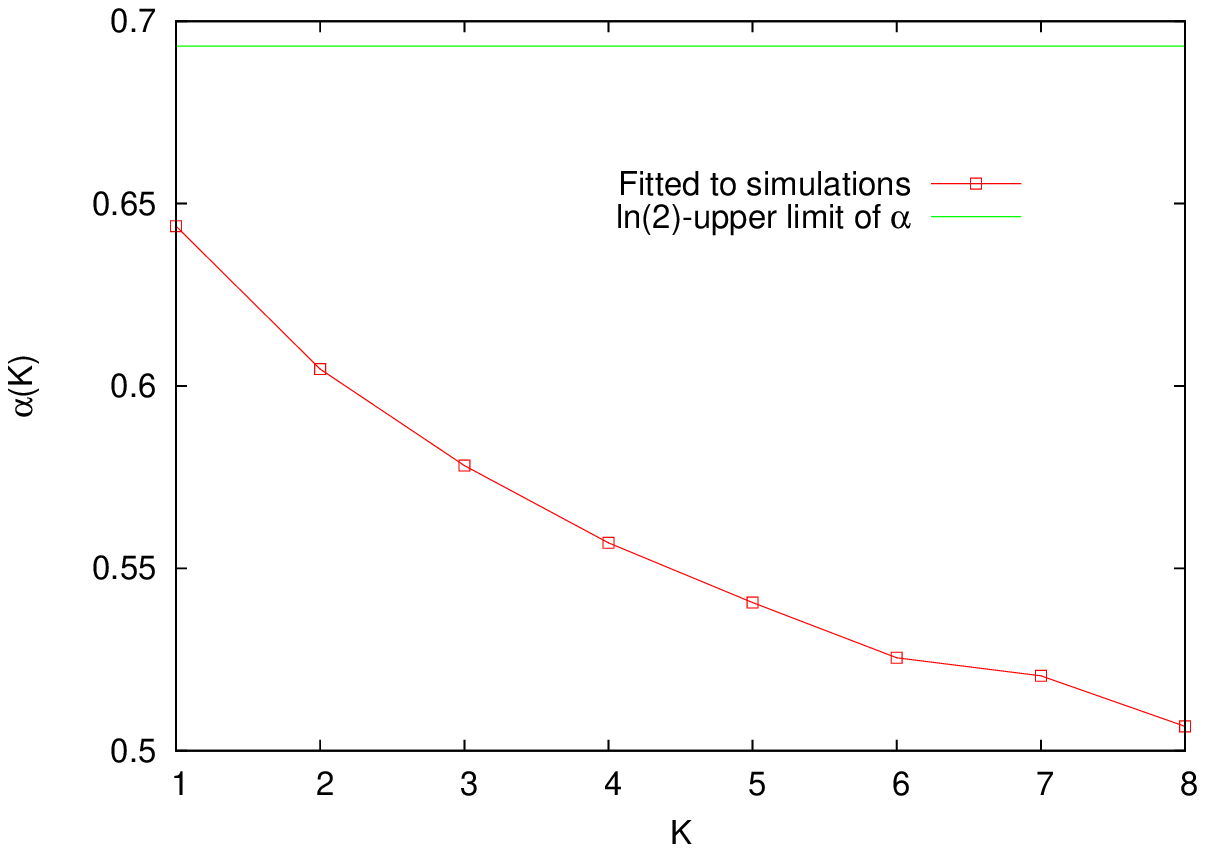}
\caption{\label{Ex_BoA_fig} (Color online) The left panel shows expected size of the
  BoA of the GM as a function of $L$ for various values of $K$ on
  semi-logarithmic scales. The growth with $L$ is exponential, as
  indicated by the straight lines. For comparison the full volume of
  state space $2^L$ is also shown. The right panel shows the
  exponential growth rate $\alpha$ of the expected BoA size as function of $K$}
\end{figure}

Figure \ref{Ex_BoA_fig} shows in the left panel the
expected size of the GM's BoA $\langle b_{GM} \rangle$ as a function of
the sequence length $L$. Clearly $\langle b_{GM}\rangle$ grows exponentially, 
$\langle b_{GM}\rangle \propto \exp(\alpha L)$
 with $\ln 2 > \alpha > 0.5$ depending on $K$ (right panel of
 fig.~\ref{Ex_BoA_fig}). Since $\alpha < \ln 2$
 the fraction of state space in the BoA of the GM appears to approach
 zero asymptotically. 
Similar to the 
results for the number of accessible paths shown in fig.~\ref{ex_paths}, we see
that $\langle b_{GM}(L)\rangle$ grows very slowly with $L$ when $L$ is
increased at fixed $L-K$. 

\section{Conclusions}
\label{sec:Conclusions}
In this paper, we investigated the evolutionary accessibility of
fitness landscapes derived from Kauffman's $LK$ model. Accessibility
of these landscapes was quantified in terms of topographic
properties that govern the adaptation in the strong
selection/weak mutation regime (accessible paths) and the greedy regime of
adaptation (basins of attraction). 
The $LK$ model allows to tune the number of
interactions between genetic loci and thus the ruggedness of the
landscape. However, it is not \emph{a priori} clear how this parameter
$K$, which governs the interaction strength, should behave as function
of sequence length $L$. Therefore three different scalings of $K$ with
$L$ were considered. 

In the case of fixed difference  $L-K=const.$, the accessible paths
were found to behave quite 
similar to the totally rugged case $K=L-1$ for large $L$. However if
the \emph{fraction} $K/L$ is kept fixed, the landscapes become
smoother in the large $L$ limit in the sense that the expected 
number of accessible paths grows (possibly even super-exponentially,
see fig. \ref{ex_paths}) and that the probability of finding a
realization without any accessible paths decays, see the right panel
of fig. \ref{p0_dif_fix}. 

\begin{figure}
  \includegraphics[width=0.7\textwidth]{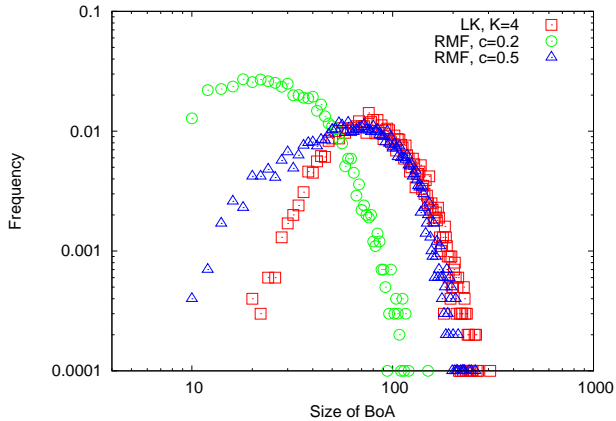}
  \caption{\label{HoC_RMF_morph} (Color online) Distribution
  of the GM's BoA for the RMF  and $LK$-models with $L=9$ and
  different values of 
  $c$ and $K$ as indicated. The value $c=0.5$ for the RMF model has
  been chosen to match the
  distribution of basin sizes as obtained from the $LK$ model. Such a match is
clearly not possible for $K=1$ and $2$, see figs. \ref{fd_BoA_K1} and
\ref{fd_BoA_K2/3}. Decreasing $c$ to the value 0.2 moves the RMF 
distribution close to the HoC distribution shown in the inset of
fig.~\ref{HoC_BoA_fig}}
\end{figure}

In the case of fixed $K$, the behavior of  the mean number of
accessible paths $\langle n_L \rangle$ was
found to be in accordance with the naive expectation that the
resulting FL should be very smooth since $K/L\to 0$ as $L \to
\infty$, see fig. \ref{ex_paths}. However the probability of no
accessible paths  $\pi_L(0)$ showed a surprising behavior. For $K\geq
3$, $\pi_L(0)$ declines monotonically with $L$ and the curves are very
similar. For $K=1$, however, $\pi_L(0)$ \emph{increases} with $L$. In the case
$K=2$, $\pi_L(0)$  seems almost independent of $L$, thus this appears
to be a marginal case separating the two regimes. This was not
expected because $K=1$ is the (non-trivial) case with the least amount
of interactions and was thus expected to be smoother than those for
larger $K$. 

This difference in behavior of $K=1$ and $K \geq 3$ is also reflected in
the full distribution $\pi_L(n)$ shown in figs. \ref{full_K=1_fig} and
\ref{full_K=2/3_fig}. In one case, $\pi_L(n)$ remains dominated by
characteristic peaks whereas for $K\geq 3$, the curves become smoother
with increasing sequence length eventually showing the behavior
found in fig.~\ref{lk_paths_full} for the HoC case $K=L-1$. 

The full distribution of the BoA of the GM under
greedy dynamics also showed a characteristic spike structure for
$K=1$, see fig. \ref{fd_BoA_K1}. As for the distribution of the number
of accessible paths,
the structure remained essentially unchanged for the sequence lengths
considered. Under rescaled variables $b/2^L$, the dominant spikes retain
their positions, which indicates that the reason for these dominant
features of the distribution is commensurate with the total number of
states $2^L$.

At this point we can only speculate about the reasons for the marked
change in behavior that seems to occur in the $LK$ model around $K=2$.
We have noted above that the $LK$-model can be seen as a dilute spin
glass with $p$-spin couplings up to $p=K+1$. In the SG context it is
well known that $p$-spin models behave qualitatively different for $p
= 2$ and $p \geq 3$, respectively, with regard to static \cite{Gross1984} as
well as dynamic \cite{Kirkpatrick1987,Bovier2008} properties. 
It remains to be seen if the observations reported here can be understood
from the SG perspective.

The RMF model, which was briefly alluded to here, is another model
which yields tunably rugged fitness landscapes. However, unlike
the $LK$ model, the distributions obtained for the RMF model show no
signs of discrete structures and, as $c \to 0$, the curves change
smoothly into those from the HoC model, see for example
fig.~\ref{HoC_RMF_morph}
and results presented in \cite{fkdk2011}. From a fitness landscape point 
of view it is remarkable that the way in which `intermediate
ruggedness' arises (by an external `field' or by internal interactions)
plays such an important role, while from a spin glass point of view, it
is perhaps surprising that the two ways of generating intermediate ruggedness
yield similar behavior at all.

\paragraph{Acknowledgments.} We would like to thank Johannes Berg,
Anton Bovier, Bernard Derrida and Remi Monasson for useful discussions
and remarks, and Alexander Kl\"ozer for his contributions in the
initial stages of this project. This work was supported by DFG within SFB 680 and BCGS. JF
acknowledges financial support from Studienstiftung des deutschen
Volkes.

\end{document}